\documentclass[aps,prl,reprint,superscriptaddress]{revtex4-1}

\usepackage{amsmath}
\usepackage{graphicx}
\usepackage{epstopdf}
\DeclareMathOperator{\sech}{sech}

\begin{document}


\title{Direct single-shot observation of millimeter wave superradiance in Rydberg-Rydberg transitions}


\author{David D. Grimes}
\author{Stephen L. Coy}
\author{Timothy J. Barnum}
\affiliation{Department of Chemistry, Massaschusetts Institute of Technology, Cambridge, Massachusetts 02139, USA}

\author{Yan Zhou}
\affiliation{JILA, National Institute of Standards and Technology and University of Colorado, Department of Physics, University of Colorado, Boulder, Colorado 80309-0440, USA}

\author{Susanne F. Yelin}
\affiliation{ITAMP, Harvard-Smithsonian Center for Astrophysics, Cambridge, Massachusetts 02138, USA}
\affiliation{Department of Physics, University of Connecticut, Storrs, Connecticut 06269, USA}

\author{Robert W. Field}
\email{rwfield@mit.edu}
\affiliation{Department of Chemistry, Massaschusetts Institute of Technology, Cambridge, Massachusetts 02139, USA}


\date{\today}

\begin{abstract}
We have directly detected millimeter wave (mm-wave) free space superradiant emission from Rydberg states ($n \sim 30$) of barium atoms in a single shot. We trigger the cooperative effects with a weak initial pulse and detect with single-shot sensitivity and 20 ps time resolution, which allows measurement and shot-by-shot analysis of the distribution of decay rates, time delays, and time-dependent frequency shifts. Cooperative line shifts and decay rates are observed that exceed values that would correspond to the Doppler width of 250 kHz by a factor of 20 and the spontaneous emission rate of 50 Hz by a factor of $10^5$. The initial superradiant output pulse is followed by evolution of the radiation-coupled many-body system toward complex long-lasting emission modes. A comparison to a mean-field theory is presented which reproduces the quantitative time-domain results, but fails to account for either the frequency-domain observations or the long-lived features. 
\end{abstract}


\maketitle

\section{}

Superradiance is an effect in which emitters radiate collectively and coherently due to constructive interference between electric dipoles that communicate with each other via a shared radiation field \cite{Dicke1954}. Subradiance is exactly the opposite - a collective inhibition of radiation due to destructive interference between the radiation from an array of dipoles \cite{Bienaime2012,Guerin2016}. These cooperative phenomena provide insights into fundamental many-body physics \cite{MacGillivray1976,Gross1982,Lin2012} and suggest applications ranging from quantum information storage \cite{Clemens2003,Black2005,Svidzinsky2010} to narrow linewidth lasers \cite{Meiser2009,Bohnet2012a,Bohnet2012,Svidzinsky2013}.\par

The large electric dipole transition moments and long wavelengths associated with Rydberg-Rydberg transitions make these transitions natural candidates for observing collective effects at relatively low atom number densities ($\rho \sim 10^6$ cm$^{-3}$). Hydrogenic scaling rules show that, for $\Delta n = 1$ (where $n$ is the principal quantum number), the transition dipole moment between Rydberg states scales as $\mu \propto n^2$ and the wavelength scales as $\lambda \propto n^3$. The transition dipole moment controls only the individual atom spontaneous decay rate, which is independent of density. Collective effects, however, scale in multiples of the spontaneous decay rate. The multiplicative factors scale as the optical depth ($OD = \rho\lambda^2L$, where $\rho$ is the density and $L$ is the length of the sample) \cite{Lin2012} or the relative density ($RD = \rho\lambda^3$) \cite{Javanainen2014}. Thus, these two scaling rules result in strong collective effects at several orders of magnitude lower densities than between valence states of atoms or molecules. Superradiant emission is also focused primarily on the transition with the smallest $\Delta n$ allowed by the $\Delta\ell = \pm 1$ angular momentum selection rule \cite{Wang2007}. For Rydberg states with $n \sim 30$, $\Delta n = 1$ transitions lie at $\sim$300 GHz ($\lambda \sim 1$mm) and have transition moments on the order of 500 debye\footnote{$1$ debye = 3.336$\cdot$10$^{-30}\:$C$\cdot$m $ = 0.393\:$\textit{e}$\cdot$\textit{a}$_0$}.\par

Typically, the total number of Rydberg state atoms in a single experiment has been too small to permit direct detection of the emitted electric field. Previous studies of collective effects in ensembles of Rydberg states have relied on state-selective field ionization detection in order to infer \textit{indirectly} that superradiance has occurred \cite{Gross1979,Moi1983,Wang2007,Han2014}. \textit{Direct} detection of the emitted electric field was achieved in cavity based maser experiments in the 1980s, but was sensitive only to the intensity of the emission, not to its frequency or phase \cite{Moi1980,Goy1983}.\par

Recent improvements in mm-wave technology \cite{Brown2008,Dian2008,Park2011,Prozument2011,Colombo2013} and atomic beam sources \cite{Maxwell2005,Patterson2009a,Zhou2015} have enabled direct observation of superradiance \textit{in a single shot}. In this letter, we report direct heterodyne detection of the time-dependent emitted electric field that arises from superradiance in a sample of Barium Rydberg atoms. Because we trigger emission in the inverted system with an initial weak pulse, which stabilizes the dynamics otherwise initiated by spontaneous emission, we are able to	 observe cooperative decay rates and line shifts of the emission frequency thousands of times larger than the natural decay rate and long-duration coherent, cooperative emission from the sample volume in a single experiment. We compare those observations to a mean-field theoretical treatment. Mean-field theory is able to describe much of the quantitative time-domain behavior of the superradiant pulse, but it fails to provide even a qualitative description of the frequency domain behavior or the nature of the long-lived emission.\par

\begin{figure}
\includegraphics[width=0.48\textwidth]{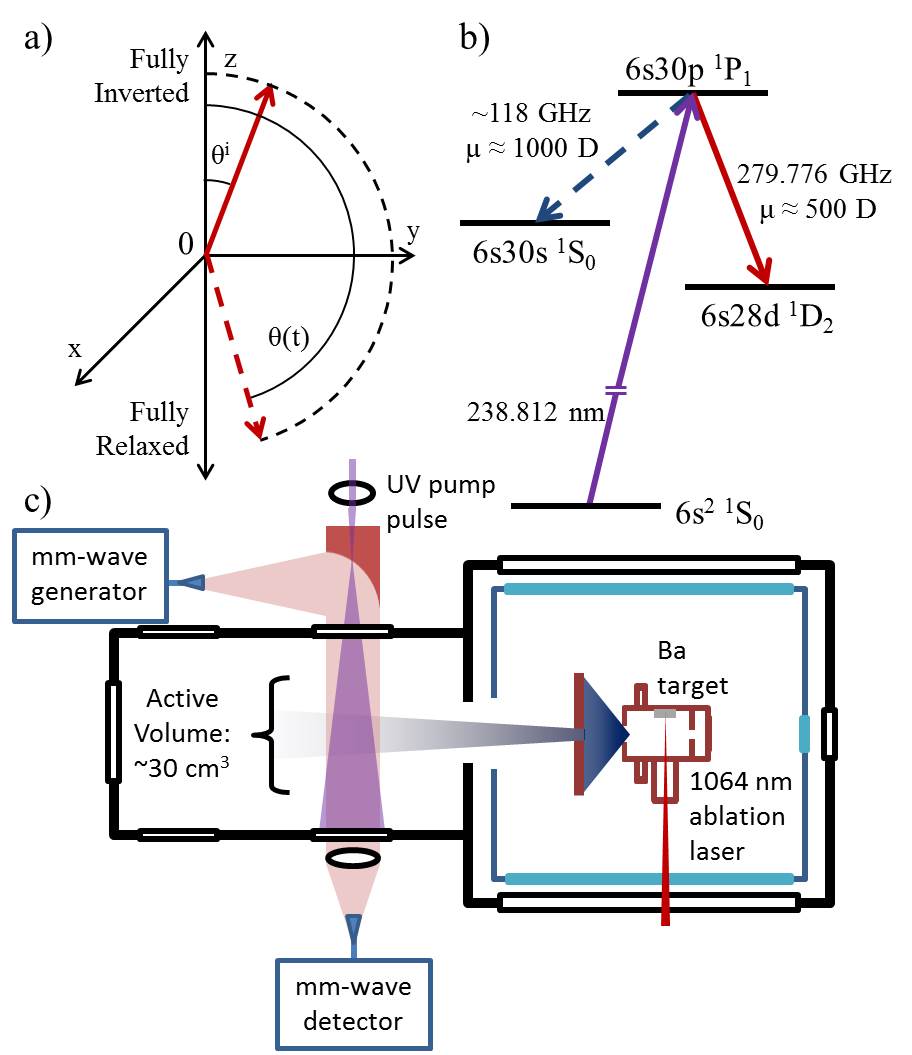}
\caption{a) The Bloch angle formalism for describing mean-field superradiance. b) Energy level diagram. The dotted arrow connecting the 6s30p and 6s30s states indicates that superradiant emission on that transition is spontaneous rather than triggered. c) Schematic diagram of the experimental setup. Details of the mm-wave generation and ablation conditions are included in the text.}
\label{fig:Picture1}
\end{figure}

As first considered by Dicke \cite{Dicke1954}, a collection of N coherently prepared two-level systems with physical separation much smaller than a wavelength can be described as a single spin-N/2 system. The evolution of this system is described classically by a vector evolving on the Bloch sphere analogous to a classical damped pendulum \cite{Gross1982}. If the system is initially inverted, it remains stationary until the first spontaneous emission event occurs or an oscillating on-resonance electric field (potentially from a blackbody emitter) is encountered. This first event tilts the Bloch vector an angle $\theta^i$ from the z-axis (see Fig. (\ref{fig:Picture1}a)), and initiates evolution according to the equations:

\begin{subequations}
\begin{gather}
\label{eq: Bloch}
\frac{d\theta}{dt} = \frac{1}{2 T_R}\sin(\theta)\\
\bar{N}(t)= \frac{1}{2}(N_{e} - N_{g}) = \frac{N}{2}\cos⁡(\theta(t))
\end{gather}
\end{subequations}

\noindent where $\theta$ is the angle between the Bloch vector and the z-axis, $\bar{N}$ is half the difference in population between the excited ($N_e$) and ground ($N_g$) states, and $T_R$ is the characteristic superradiance time given by $T_R=(8\pi)/(3 \rho \lambda^2 L A_{21})$, where $A_{21}$ is the Einstein A coefficient for the transition. The corresponding radiated field magnitude is:

\begin{equation}
I(t)=-\hbar\omega_0 \frac{d\bar{N}}{dt}=\frac{\hbar\omega_0 N}{2T_R} \sech^2⁡\left[\frac{1}{2T_R}(t+T_D)\right]
\label{eq: Int}
\end{equation}

\noindent with $T_D$ the characteristic delay time given by $T_D = 2T_R \log(\theta^i/2)$ and $\hbar\omega_0$ the energy difference between the two states. However, these equations are correct only for a sample length much smaller than a wavelength \cite{Moi1983} and therefore Eq. (\ref{eq: Int}) cannot be correct for our experimental conditions. For active volumes larger than $\lambda^3$, the spatial variation across the sample in the magnitude of the shared electric field causes Eqs. (1) to take the form:

\begin{subequations}
\begin{gather}
\frac{d\theta}{dt} = \frac{1}{2 T_R}J_1(\theta)\\
\bar{N}(t)=\frac{N}{2}J_0(\theta(t))
\label{eq: LgSample}
\end{gather}
\end{subequations}

\noindent where $J_0$ and $J_1$ are the zeroth and first order Bessel functions. The population difference no longer decreases to $-N/2$, as in the small sample case, but rather $\bar{N}(t\rightarrow\infty)\approx-N/10$ due to dephasing between different regions of the active volume. The result is that $\sim 40\%$ of the initially excited atoms are left in the excited state. In mean field theory, the remaining excited state atoms do not radiate at all. In reality, these atoms dephase and radiate due to Doppler broadening and dipole-dipole interactions.\par

The radiated field magnitude can be calculated as in Eq. (\ref{eq: Int}), but can no longer be expressed analytically. However, the hyperbolic secant lineshape form from Eq. (\ref{eq: Int}) remains a good approximation of the magnitude. The primary difference is that the characteristic evolution time changes from $2T_R$ in Eq. (\ref{eq: Int}) to $4T_R$ in the large-sample case, because the evolution stops before all emitters have returned the ground state \cite{Moi1983}.\par

We generate an atomic beam of barium atoms using a neon buffer gas cooled atomic beam similar to that described in reference \cite{Zhou2015} and summarized briefly here. Barium atoms are generated by ablation of a metal target inside a buffer gas cell with 50 mJ/pulse of the 1064 nm fundamental of a Q-switched Nd:YAG laser focused to a 1 mm$^2$ spot size. The ablation plume of Ba atoms is entrained in a 20 SCCM (standard cubic centimeters per minute) flow of 20 K neon ($\sim10$ mtorr pressure inside the cell) and is hydrodynamically expanded into vacuum and then through a 2 cm diameter skimmer held at 6 K to form a loosely collimated atomic beam. This beam is crossed by the laser and mm-wave pulses in a separate chamber 15 cm downstream. This atomic beam has a lab frame velocity of 180 m/s, transverse translational temperature of 5 K, and transverse Doppler width of 250 kHz.\par

Atoms are excited into Rydberg states by pumping the 6s$^2$ $^1$S$_0$ $\rightarrow$ 6s30p $^1$P$_1$ transition with a 238.812 nm 5 ns laser pulse produced by the doubled output of a seeded Nd:YAG pumped dye laser. This typically excites $3*10^8 - 1.5*10^9$ total atoms into a single Rydberg state in a volume of 30 cm$^3$ ($\rho = 1*10^7 - 5*10^7 $cm$^-3$) with characteristic length of 15 cm. Immediately following excitation to the Rydberg state, a 10 ns mm-wave pulse, on resonance with the 6s30p $^1$P$_1$ $\rightarrow$ 6s28d $^1$D$_2$ ($A_{21} \approx 50$ Hz) transition at 279.776 GHz, triggers the superradiant emission (with the $^1$P$_1$ state as the excited state and the $^1$D$_2$ state as the lower state). This pulse is formed using a Virginia Diodes Active Multiplier Chain (AMC) to multiply the frequency output of a 12 GS/s Agilent Arbitrary Waveform Generator (AWG) mixed with a fixed frequency 8.8 GHz local oscillator (LO). The mm-wave pulse energy is 300 pJ, which corresponds to an $\theta^i = \pi/40$ initial tip angle of the Bloch vector. The time delay betweem excitation and initiation is much shorter than the typical time required for a spontaneous emission event to occur or for an on-resonance blackbody photon to interact with the system. We take the tip angle to be determined entirely by the mm-wave pulse. We detect the subsequent superradiant pulse at 279.776 GHz directly in the time domain by heterodyning against a LO set to 277.2 GHz. This radiation is generated by the same AWG and 8.8 GHz LO, using a second AMC for multiplication. The resultant output is recorded on a 50 GS/s oscilloscope.\par

A schematic level diagram and representation of the experiment are shown in Figs. (\ref{fig:Picture1}b) and (\ref{fig:Picture1}c). The data discussed here is from 111 individual shots, of which 56 had a signal to noise ratio large enough to be fitted. The largest effect on the signal to noise ratio of each shot was the shot-to-shot variation in Rydberg density that results from variations in both the ablation laser intensity and the dye laser intensity.\par

\begin{figure}
\includegraphics[width=0.48\textwidth]{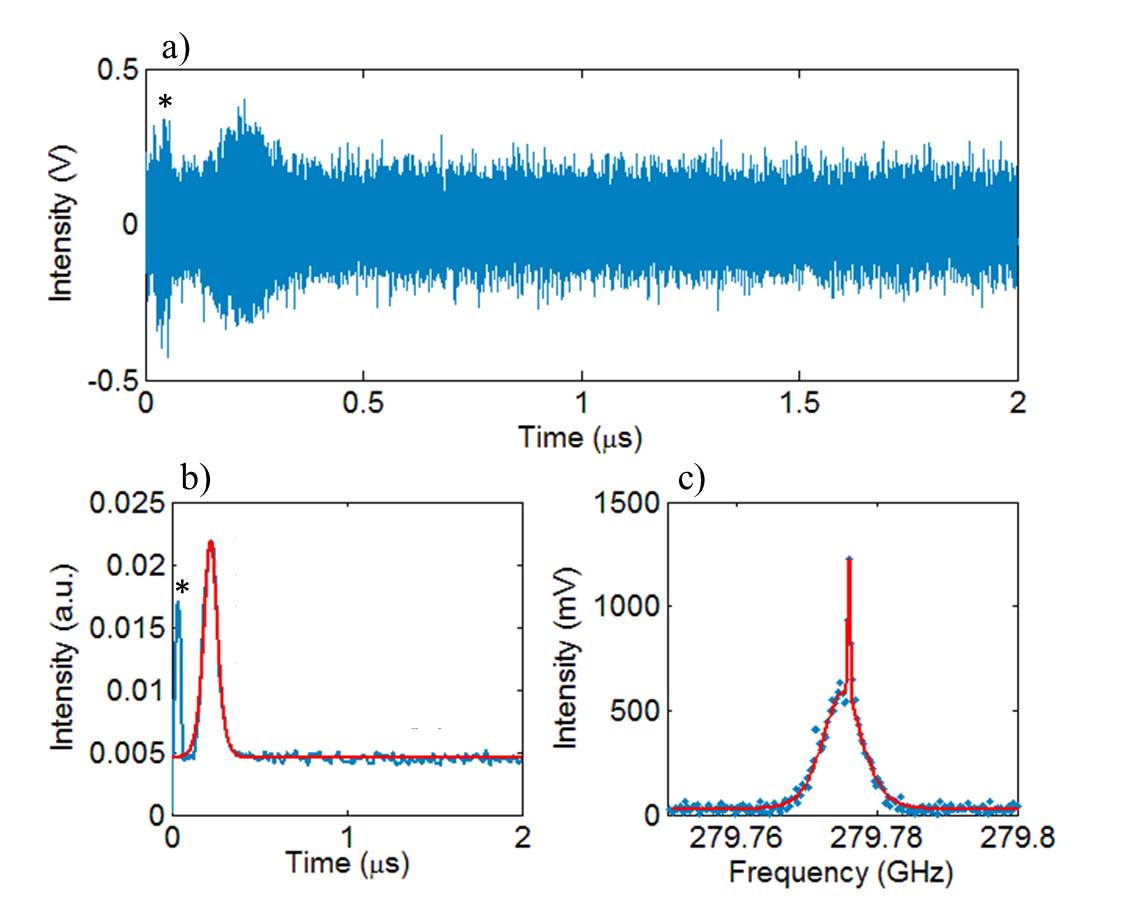}
\caption{a) Raw single-shot data trace recorded in the experiment. The starred feature is the tipping pulse that initiates the superradiance, and the larger feature is the superradiant emission. b) The digitally filtered electric field intensity profile is shown in blue, and the fit to the mean-field emission functional form is shown in red. The starred feature is again the tipping pulse that triggers the superradiance. c) The Fourier transform of the raw data shown in part a. The blue points are the data and the red line is the fit to a sum of the two lineshape functions, as described in the text.}
\label{fig:Picture2}
\end{figure}

An example of the raw data from the oscilloscope is shown in Fig. (\ref{fig:Picture2}a). The initial, starred peak is the initial mm-wave tipping pulse, while the second, larger, and broader peak is the superradiant emission. Due to our large detection bandwidth, the vast majority of the noise is at frequencies far from the resonance frequency, thus we make use of digital filtering methods to improve our signal to noise. Briefly, we independently measure the low-density resonance frequency (279.776 GHz) and multiply the signal by a sine and a cosine wave at that frequency to extract the in-phase and quadrature components of the signal. We then use a phase-conserving 10 MHz low pass filter to remove the high frequency noise. We calculate the time-dependent radiated field magnitude and phase independently:

\begin{subequations}
\begin{gather}
I(t)=s^2+c^2 \\
\phi(t)=\tan^{-1}⁡(s/c)
\label{eq: Quad}
\end{gather}
\end{subequations}

\noindent where $s$ and $c$ are the in-phase and quadrature components of the signal, and the phase is calculated using a four quadrant arctangent function. After filtering, we fit the time-domain field magnitude to Eq. (\ref{eq: Int}). An example of a fitted filtered signal is shown in Fig. (\ref{fig:Picture2}b). The width and delay of the signal ($T_R$ and $T_D$) are fitted separately and match the expected relationship given earlier for an initial tip angle of $\pi/40$.\par

Figure (\ref{fig:Picture2}c) shows the Fourier transform of the signal shown in Fig. (\ref{fig:Picture2}a), gated to exclude the initial tipping pulse, and a fitted lineshape. Two features are prominent. First, the broad feature is the signal associated with the superradiant burst of radiation. It has a width of 7 MHz and is shifted in frequency 4 MHz lower than the transition frequency at low density. The remaining narrow feature is the signal remaining after the superradiant evolution concludes, with width ($\sim$250 kHz) consistent with Doppler broadening and with no observable shift from the transition frequency at low density. This signal could arise from either subradiance modes of the sample \cite{Bienaime2012,Guerin2016} or radiation trapping \cite{Holstein1947,Fleischhauer1999}. However, our detection system is only sensitive to coherent radiation, suggesting that this signal is due to subradiance.\par
 
The fitted lineshape is the sum of a narrow Gaussian peak, centered at the low-density resonance and a hyperbolic secant peak, the center frequency of which was allowed to vary. When performing a Fourier transform with a gate that excludes both the initial tipping pulse and the superradiant pulse, only the narrow feature remained.  Intentionally attenuating the pump laser leads to a reduction of both signals, implying that the narrow feature is fully cooperative in nature, not due to low density portions of the sample.\par

\begin{figure}
\includegraphics[width=0.4\textwidth]{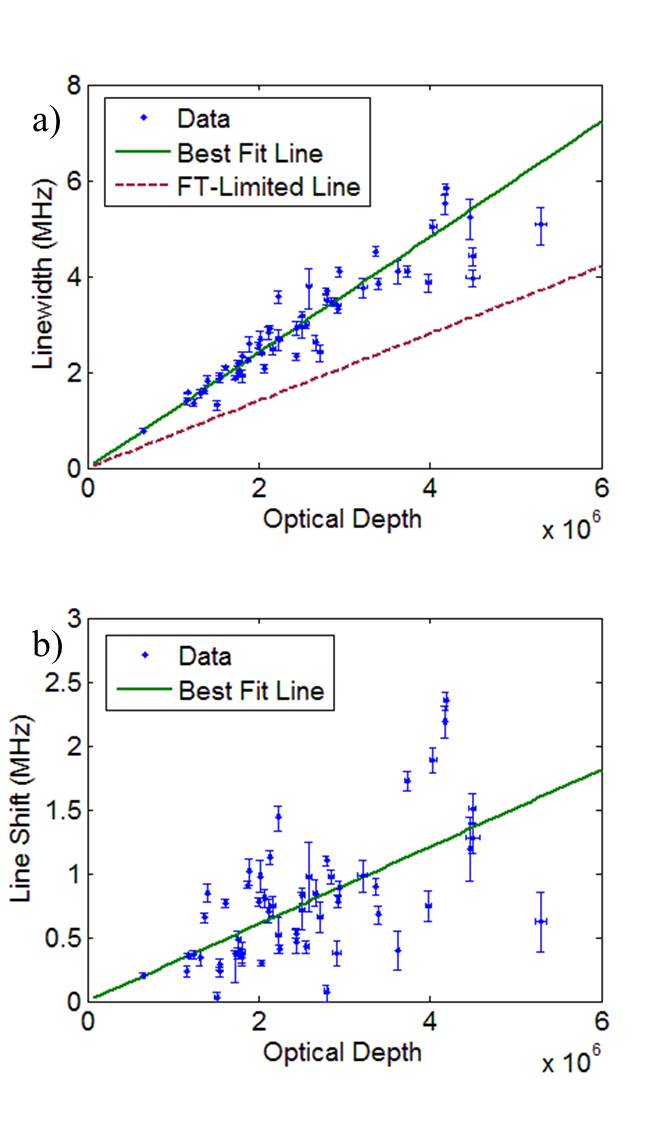}
\caption{a) Relationship between the optical depth of a superradiant sample and the linewidth of the emitted radiation. Blue points are the data, the green line is the best fit to the data, and the red dashed line is the expected linewidth if the emission were Fourier transform limited. Error bars represent 95$\%$ confidence intervals. b) Relationship between the optical depth of a superradiant sample and the line shift. Blue points are the data and the green line is the best fit to the data. Error bars represent 95$\%$ confidence intervals.}
\label{fig:Picture3}
\end{figure}

The shot-to-shot variation in number of excited atoms allows for immediate shot-by-shot investigation of the dependence on optical depth determined from time domain fits of $T_R$ using Eq. (\ref{eq: Int}). The width of the superradiant feature varies linearly (R$^2$ = 0.87) with optical depth, as shown in Fig. (\ref{fig:Picture3}a). This is expected, as $T_R$ is inversely proportional to optical depth. However, the width is consistently larger than the Fourier transform limited linewidth associated with a time domain hyperbolic secant signal with characteristic width $T_R$. In Fig. (\ref{fig:Picture3}a), the solid green line shows a linear fit of the optical depth vs. linewidth, while the dashed red line shows what the width would be if it were Fourier transform limited. The observed excess frequency width implies that the frequency chirps across the emission feature.\par
	
\begin{figure}
\includegraphics[width=0.48\textwidth]{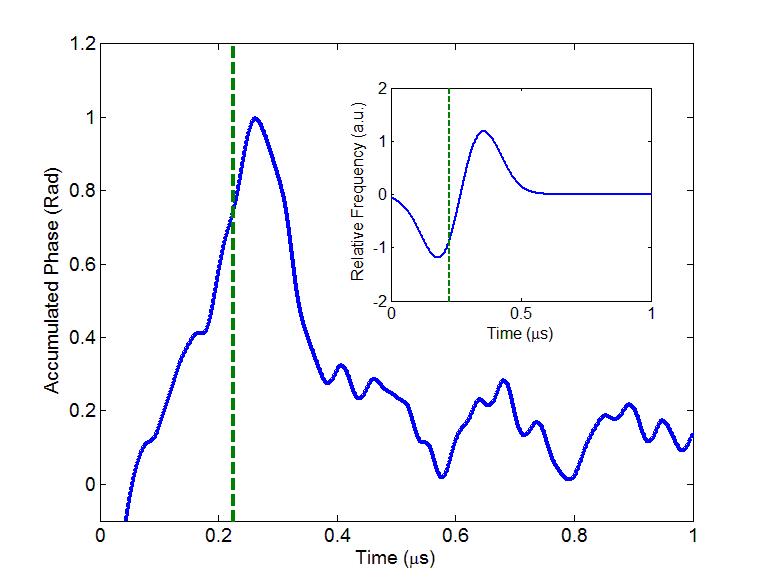}
\caption{The recorded phase as a function of time, obtained by demodulation at the low-density resonance frequency. The green dashed line indicates the time at which the electric field magnitude maximum occurs. Positive slopes indicate a frequency that is shifted lower than the resonance frequency, while negative slopes indicate a shift to higher than the resonance frequency. The inset shows a schematic frequency evolution associated with the displayed phase evolution, as explained in the text, and the green dashed line indicates the time at which the field magnitude reaches its maximum value. The low-density resonance frequency is taken as 0.}
\label{fig:Picture4}
\end{figure}

The frequency at any time point can be recovered through the derivative of the phase, which is sampled by our methods of detection and filtering. The blue solid trace in Fig. (\ref{fig:Picture4}) shows the phase evolution of the signal shown in Fig. (\ref{fig:Picture2}a). In the absence of a reliable model, the phase evolution was fit to a series of lineshape functions (Gaussian, Lorentzian, hyperbolic secant) and the derivative of each was taken in order to determine the frequency as a function of time. Qualitatively, each fit model produced the same results; the frequency evolution determined from the Gaussian fit is shown in the inset to Fig. (\ref{fig:Picture4}), plotted relative to the low density emission frequency. Of note is that the frequency is \textit{chirping during the emission}, and that the time of the maximum in the phase evolution, when the frequency crosses through the low-density resonance frequency, does not coincide with the maximum of the field magnitude in the time domain, which is indicated by a green dashed line. Mean field theory predicts that the magnitude maximum and the passage of the frequency through resonance should occur at the same time. This mismatch also causes the frequency shift observed in the frequency domain, as the emission is most intense while the frequency is shifted away from the low-density value, despite the fact that the chirp appears symmetric around the low-density value. As a comparison, in a quantum many-body treatment, the chirp is not quite symmetric because the broadening is not Gaussian, but it does predict an overall shift to lower frequencies \cite{Lin2012}.\par

The relationship between Rydberg optical depth and the observed frequency shift could not be documented in this work. The 6s30s $^1$S$_0$ state lies $\sim$4 cm$^{-1}$ below the 6s30p $^1$P$_1$ state, as shown in Fig. (\ref{fig:Picture1}b), and the 6s30p state can superradiantly decay to it \textit{without} a trigger pulse. This population decay is triggered by either a spontaneous emission event or on-resonance black body radiation, and begins both at a random location in the atomic sample and at an uncontrolled time in the experiment. Therefore, the distribution of emitters taking part in the triggered superradiance changes on a shot-by-shot basis, as do all propagation effects, making it impossible to establish a quantitative relationship between number density, geometry, and frequency shift. However, there is a weak dependence (R$^2$ = 0.38) of frequency shift on optical depth, as shown in Fig. (\ref{fig:Picture3}b).\par

To conclude, we have directly observed mm-wave superradiant emission between Rydberg states, and modeled much of the quantitative time-dependent evolution of the field magnitude in a mean-field model.  Additionally, we have recorded the time-dependent phase and frequency responses of our highly cooperative system, which show both a frequency \textit{chirp} and an overall frequency \textit{shift} of the superradiant signal by $\sim10^5$ times the natural linewidth, or $\sim20$ times the Doppler linewidth.  If shot-to-shot geometry variations could be controlled, by exciting directly to a 6sns $^1$S$_0$ Rydberg state which predominantly decays to a single 6sn'p $^1$P$_1$ state, uncontrolled competing effects would be minimized and both density dependent frequency shifts and the evolution into long-lived, potentially subradiant emission modes could be investigated.

\subsection{}
\subsubsection{}

\begin{acknowledgments}
We thank Profs. Edward Eyler and John Muenter for valuable comments and feedback.  This material is based upon work supported by the National Science Foundation under Grant No.  CHE-1361865.  David D. Grimes was supported by the Department of Defense (DoD) through the National Defense Science $\&$ Engineering Graduate Fellowship (NDSEG) Program.

\end{acknowledgments}

\bibliography{Superradiance_PRL}

\end{document}